\documentclass[11pt]{article}
\usepackage{amsmath}
\usepackage{amssymb}
\usepackage{amscd}
\usepackage{graphicx}
\usepackage{epstopdf}
\usepackage{wrapfig}
\title{Nature's software}
\date{\small{23 January 2011} }
\author{Daniel Canarutto\\[6pt]
{\small\it Dipartimento di Matematica Applicata ``G. Sansone'', }\\
{\small\it Via S. Marta 3, 50139 Firenze, Italia}\\
{\small email:~{\tt{daniel.canarutto@unifi.it}}}\\
{\small {\tt{http://www.dma.unifi.it/\char126 canarutto}}}}
\begin{document}
\bibliographystyle{alpha}
\maketitle
\begin{abstract}\noindent
I bring forward some arguments to support the thesis that nature
is fundamentally discrete,
and present my own thoughts about the direction in which one could look
for a possible, consistent ``theory of everything''
describing gravitation and quantum particles.
\end{abstract}

\section*{Introduction: why the reals?}

According to one point of view, particles are an ``epiphenomenon'' and
the fundamental theory is one of fields.
Others think that the foundations of physics are discrete,
and that the continuous description is just a useful tool at some level.
The former approach is fairly standard in theoretical physics,
where one works out the ``quantization'' of classical fields.\footnote{
See for example the introduction of S.~Weinberg's
\emph{The Quantum Theory of Fields},
Cambridge University press (1996).} 
On the other hand,
various proposals going in the latter direction have appeared even recently
and are actively researched.\footnote{
\label{footnote:discretethreads}
For a sample of some of the threads see e.g.:\\
E.P.~Verlinde,
`On the Origin of Gravity and the Laws of Newton',
arXiv:1001.0785v1;\\
C.~Rovelli,
`Loop quantum gravity: the first twenty five years',
arXiv:1012.4707v3;\\
R.D.~Sorkin, `Causal Sets: Discrete Gravity',
arXiv:gr-qc/0309009v1.
} 
So, the issue Niels Bohr thought he settled with his complementarity principle
actually continues to stir debates.

Now the discrete aspects in the physics of quantum particles are evident,
so a theory founded on continuous concepts has the problem of explaining
how discreteness actually arises;
tentative arguments take a hint from the ``vibration modes''
of a bounded system (e.g.\ a string),
but, as far as I know,
no definitely convincing results have been obtained.

Furthermore,
we may note that saying that the foundations of physics are continuous
implies seeing the real numbers as a primitive concept.
To me, this seems disputable.
The field $\mathbb{R}$ of reals is a highly abstract construction.
Its elements are defined to be certain equivalence classes of infinite sequences,
and have many non-intuitive properties.
Most reals cannot be characterized by finite information
(the real numbers we deal with in practice constitute a very small subset).
The reason why they are so important is that in $\mathbb{R}$
we can introduce a suitable notion of limit,
and prove all those theorems which make calculus consistent.
In other terms, the reals provide us with a powerful and sound context
for calculations.
This is not the same as saying that they are to be included
into the fundamental notions of physics.
I'd rather say that taking the reals for granted,
and including differentiable manifolds in the basic setting of a theory,
amounts to starting from strong, involved assumptions.

I'm not convinced by the contrary argument
that the real numbers are essential for calculations,
and so should be included in the fundamental assumptions.
This idea, that calculability can't be separated from the basic assumptions,
is actually widespread in theoretical physics;
on the other hand one may contend that a conceptually clearer theory,
based on simpler assumptions,
may well require harder calculations
(this point was explicitly brought forward by Einstein himself).

\section*{What is a quantum particle?}\label{s:What is a quantum particle?}

Several discussions of basic aspects of Quantum Mechanics
concern experiments in which
a particle source and a screen (the detector)
are separated by polarizing filters, or by a wall with slits, or whatever.
In Young's experiment the wall has two slits,
and in a classical context one would say that each particle
passes through one slit.
Adjusting the source to be dimmer and dimmer,
the screen detects single particles;
the time separation between two conscutive flashes can be rendered
large at will,
but eventually the distribution of the flashes sums up
to the same interference pattern determined by the bright source
(that pattern is destroyed if one tries to detect the particles at the slits).

So we detect certain quantum events, the absorption of single particles,
and we have good reasons to assume that each event is correlated
to an event occuring at the source, the emission of one particle.
It's then natural to view the particle \emph{exactly}
as this correlation between two observed events.
Though our (classical) mental habits would suggest that `something'
has traveled from the first event to the second,
this is not the right description at a more fundamental level.
The two events and their (abstract) correlation are just everything
that can be regarded as ``real''.

Now suppose that some external Observer looks at our universe,
not being subjected to our universe's time,
and sees the whole spacetime at one glance
(the kosmos from the beginning to the end).
This object looks to the Observer as a
(for us,  huge and extremely complex)
network of correlated events.
Considering a particular photon,
the Observer might see that a certain event,
occured at recombination era 400,000 years after the big bang,
is correlated with an event occured at a radiotelescope on Earth
14 billion years after the big bang.
He might say that a photon was emitted at the recombination era
and detected 14 billion years later,
but \emph{not} that something was wandering all this time through the cosmos,
waiting to be detected.
For the Observer, only events and their correlations exist.

We humans can't observe the kosmos in the same way
but may have, at some given time, an incomplete knowledge.
We may have detected an emission event, or know that it has occurred,
but could not spot (at least, not yet) the correlated absorption event,
which however \emph{must} occur somewhere, at some time.
This does not mean that we can say nothing about it,
actually we can calculate \emph{quantum probabilities}:
knowing about the emission event allows one to calculate the probability
that the absorption event takes place in a certain position at a given time.

\section*{Spacetime and gravitation}\label{s:Spacetime and gravitation}

The network of correlated events,
where each correlation is called a ``particle'',
constitutes a discrete structure;
now, rather than assume that it is contained in something else,
we could view it as the fundamental reality,
at least as far as our investigation can go.
Perhaps there exists no underlying continuum,
and spacetime and its metric arise as mathematical
notions which are suitable for describing the universe on macroscopic scales.
They might be of a statistical nature, analogous to thermodynamical functions.
Thus a consistent formulation of the fundations of physics
could be achieved not requiring a unification in the usual sense
that all interactions should be on the same footing.
In particular, gravitation could be seen as a kind of ``residual'' force,
emerging in the macroscopic description
and not mediated by any particle.
Though this sounds similar to Verlinde's ideas,\footnote{
Quoted in footnote~\ref{footnote:discretethreads}.} 
actually I see it as a more radical ansatz,
involving the complete elimination of the geometry
from fundamental notions.
It's also, in my opinion,
the most possible \emph{relational} ansatz.\footnote{
For a discussion of this concept see for example
C.~Rovelli,
`Half way through the woods',
Lecture presented at the `34th Annual Lecture Series'
of the Center for the Philosophy of Science
of the University of Pittsburgh (1994).} 
As far as I know, the closest ideas to what I'm trying to convey,
in the literature, have been expressed by Penrose.\footnote{
Penrose, R.:
`Angular momentum: an approach to combinatorial space-time',
in \emph{Quantum Theory and Beyond---%
essays and discussions arising from a colloquium},
Bastin T.\ editor,
Cambridge Univ.\ Press, Cambridge (1971), 151--180.
} 
Let me quote a few sentences of his:
\begin{center}
\begin{minipage}{300pt}
``The idea is to concentrate only on things which \emph{are} discrete in
existing theory and try and use them as \emph{primary concepts}---then to build
up other things using these primary concepts as the basic building blocks.
Continuous concepts could emerge in a limit,
when we take more and more complicate systems.
\par\noindent...\par\noindent
The central idea is that \emph{the system defines the geometry}...
The notion of space comes out as a \emph{convenience} at the end.''
\end{minipage}
\end{center}

Though several developments---%
Regge calculus, the theory of spin networks, loop gravity---%
are in some way related to those ideas,
I'm not aware that that program has been carried on to the point
of obtaining definite results supporting such clear-cut positions.
Thus I'd like to speculate a little about how
one could try and exploit the above arguments
in order to construct a true theory.

\begin{wrapfigure}{l}{3.7cm}
\begin{picture}(100,80)
\put(0,-15){\includegraphics{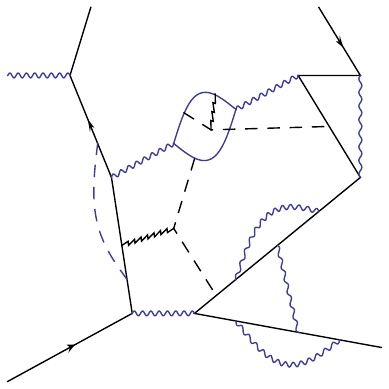}}
\end{picture}
\end{wrapfigure}\noindent
Consider a chunk of the network of events,
and draw the correlations as lines of various types.
Well, I know that this resembles a Feynman diagram,
and that such diagrams are to be considered just helps for calculations,
having nothing to do with reality;
however I'm not going to treat it really as a Feynman diagram,
and certainly I'm not saying that the lines represent paths in some space,
so please follow me for the moment.
The lines here bear various types of labels,
but none is related to spacetime in any way.
Let's say that the labels refer to an ``internal'' structure.
Since this diagram is only a piece of the big structure,
it has ``external legs''.
Our fundamental problem could then be formulated as follows:
given a large such network,
find out how the internal structure may yield
some kind of geometric relations among the external legs.
These relations won't be exact,
but only determined up to some degree of precision,
which we expect to be bigger the larger the considered network.
In practice we'll have to study possible ways,
compatible with the internal structure,
of immersing the network into some hypothetic geometry;
note that this immersion can be (partially) determined
only up to certain transformations of the geometry
(only the geometrical relations among the external legs have physical meaning).
Furthermore, note that we don't need to leave
this hypothetic large scale geometry completely undetermined;
actually we may choose one as part of our theoretical assumptions,
then one task of our study will be to understand how reasonable our choice was
(by the  way, this point is deeply related to the philosphical question
of the modalities of knowledge, which I won't examine here).

Now let me quote Penrose again:
\begin{center}
\begin{minipage}{300pt}
``The most obvious physical concept that one has to start with,
where quantum mechanics says something is discrete,
and which is connected to the structure of space-time in a very intimate way,
is in \emph{angular momentum}.''
\end{minipage}
\end{center}
Actually spin, or ``intrinsic angular momentum'',
is perhaps the most fundamental aspect of the internal structure of particles;
its intimate and subtle relation to spacetime geometry
must have a deep explanation.\footnote{
Part of my own mathematical work deals with such matters,
e.g.\ see Acta Appl.\ Math.\ {\bf 62} N.2 (2000), 187--224,
or my more recent papers posted on the arxiv.org website.} 
Now I suggest that, rather than dealing with arbitrary spin networks,
we could get to the core of the question by restricting our attention
to a simplified situation in which spin can only take the values $1/2$ and $1$;
this encompasses electrodynamics, which
is essentially the source of almost all everyday physics, classical and quantum
(excluding gravity, which we aim at treating on a different footing,
and nuclear physics).

\begin{wrapfigure}{l}{3.7cm}
\begin{picture}(100,80)
\put(0,-15){\includegraphics{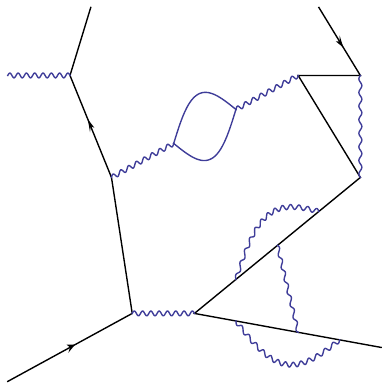}}
\end{picture}
\end{wrapfigure}\noindent
A chunk of the network of events, with the lines indicating correlation
stripped of all the internal structure except what is needed
for the present argument,
could appear as illustrated here.
Wavy lines and straight lines represent (as in usual Feynman diagrams)
photons (spin $1$) and fermions (spin $1/2$), respectively;
the arrows indicate the fermions' charge sign;
different fermion masses may be allowed.
Then, armed with one's knowledge of the mathematical relation
between spin and the Lorentz metric of classical spacetime,
one can study possible correspondences of the legs
with vectors in Minkowski space;
possibly we might eventually find that a curved structure is needed
for better fitting.
This aspect could be related to Regge calculus, 
but note that I propose that the particles themselves be the network's edges.

As for the calculation of quantum probabilities,
at least one basic question could be formulated not so differently
from standard perturbative high energy physics:
given a set of ``external legs'',
try to associate a probability to it
by taking into account (at least in principle)
all possible networks having those external legs.

I know no details
of how the above cues could be followed in practice, of course
(otherwise I'd have already published them).
Let's suppose, however, that the scheme works more or less as imagined,
and that, quoting Penrose a further time,
we succeeded in ``building up both space-time and quantum mechanics
simultaneously---from combinatorial principles'';
then we'd have reached the goal of
a peaceful cohabitation of quantum and gravitational physics.
Note, however, that neither is actually a fundamental theory in this ansatz.
Perhaps, rather than going on by building on existing theories
by adding more and more complicate features,
we could try and change our point of view of what is really fundamental.

By the way, I see this attitude as an expression of Ockham's principle,
``entities should not be multiplied beyond necessity'',
which today seems to be more often forgotten than not.

\section*{Conclusions and futher speculations: software and hardware}

Imagine a very powerful computer,
where a simulation of some virtual world was implemented
from basic entities.
Certain rules of behaviour (call them ``physical laws'')
have been assigned to these entities (``elementary particles'').
Eventually the simulation grows so complex that intelligent beings
are born in this world,
and begin to wonder about the ultimate nature of it.
How far can they reach in their quest?
They could be so smart as to guess the ``physical laws'',
namely to understand the \emph{software} of the simulation;
but the knowledge of the hardware and of the programmer
will be forever unattainable for them
(unless the programmer decides to insert some special communication).

Well, this is utter speculation,
hardly bearing any scientific weight;
but should we find that the foundations of physics are actually discrete,
then we'd be quite naturally led to see them as some kind of software.

\end{document}